\documentclass[10pt,a4paper,superscriptaddress,aps,prd,showkeys,showpacs,nofootinbib,reprint]{revtex4-2}
\usepackage[utf8]{inputenc}
\usepackage{verbatim}
\usepackage{amsmath,amsfonts,amssymb}
\usepackage[breaklinks,colorlinks]{hyperref}
\usepackage{natbib}
\usepackage{tikz}
\usepackage{float}
\usepackage{graphicx}
\usepackage{epsfig}
\usepackage{adjustbox}
\usepackage{tabularx}
\usepackage{amsbsy}
\usepackage{braket}
\usepackage[singlelinecheck=off,justification=Justified,font=small]{caption}
\usepackage{subcaption}
\hypersetup{%
,urlcolor=blue
,citecolor=blue
,linkcolor=blue
}

\newcommand\half{\textstyle\frac{1}{2}}

\begin{document}

\title{Spontaneous Hopf fibration in the two-Higgs-doublet model}
 
\author{R.~A.~Battye}
\email[]{richard.battye@manchester.ac.uk}
\affiliation{%
Jodrell Bank Centre for Astrophysics, Department of Physics and Astronomy, University of Manchester, Manchester, M13 9PL, U.K.
}

\author{S.~J.~Cotterill}
\email[]{steven.cotterill@manchester.ac.uk}
\affiliation{%
Jodrell Bank Centre for Astrophysics, Department of Physics and Astronomy, University of Manchester, Manchester, M13 9PL, U.K.
}

\label{firstpage}

\date{\today}

\begin{abstract}
We show that energetic considerations enforce a Hopf fibration of the Standard Model topology within the 2HDM whose potential has either an $SO(3)$ or $U(1)$ Higgs-family symmetry. This can lead to monopole and vortex solutions. We find these solutions, characterise their basic properties and demonstrate the nature of the fibration along with the connection to Nambu's monopole solution. We point out that breaking of the $U(1)_{\rm EM}$ in the core of the defect can be a feature which leads to a non-zero photon mass there.
\end{abstract}

\keywords{Two Higgs Doublet Model, Topological Defects}

\maketitle

{\it Introduction:} The vacuum manifold of the Standard Model of Particle Physics is a 3-sphere $({\cal M}=S^3)$ implying that there are no stable topological configurations in 3D since $\pi_2({\cal M})=I$, where $\pi_n({\cal M})$ is the $n$-th homotopy group of the manifold. However, there are interesting topological solutions with one unstable mode in 3D (sphalerons~\cite{Klinkhamer:1984di}) and 2D (electroweak vortices~\cite{Vachaspati:1992jk,Achucarro:1999it}) characterised by $\pi_3({\cal M})=Z$. Nambu suggested a monopole-like configuration~\cite{NAMBU1977505} which can be understood via a local Hopf fibration $S^3\cong S^2\times S^1$~\cite{Gibbons:1992gt}, which we discuss in more detail in \cite{supplementary}, such that $\pi_2(S^2\times S^1) = Z$. However, this configuration is known to be unstable and, if it were to be realised, it would need to be combined with a string to form so-called ``dumbbell'' configurations~\cite{Lazarides:2021bzg,Patel:2023sfm}. The dynamics of such configurations has been linked with the production of primordial magnetic fields~\cite{Vachaspati:2020blt,Patel2022} and monopoles are being actively searched for in laboratory experiments~\cite{MoEDAL:2021vix}.

We will discuss the two-Higgs Doublet Model (2HDM)~\cite{Branco2012}; in particular one where there are accidental symmetries. If the two doublets are $\Phi_1$ and $\Phi_2$ with $\Phi^T=\begin{pmatrix}\Phi_1 &\Phi_2\end{pmatrix}$ then the potential is given by $V(\Phi_1,\Phi_2)=V_2+V_4$ where $V_2=-\mu_1^2|\Phi_1|^2-\mu_2^2|\Phi_2|^2$ and $V_4=\lambda_1|\Phi_1|^4+\lambda_2|\Phi_2|^4+\lambda_3|\Phi_1|^2|\Phi_2|^2+\lambda_4|\Phi_2^{\dag}\Phi_1|^2$
which is $U(1)_{\rm PQ}$ symmetric. The symmetry can be enhanced to $SO(3)_{\rm HF}$ if $\mu_2=\mu_1$, $\lambda_2=\lambda_1$ and $\lambda_4=2\lambda_1-\lambda_3$\cite{PhysRevD.79.116004,Brawn2011}.

The particle spectrum is well understood, for example~\cite{Pilaftsis1999,Branco2012}. There are 5 Higgs particles: two of which are CP even with masses $M_h$ and $M_H$, a CP odd pseudo-scalar with $M_A$ and two charged Higgs particles with $M_{H_{\pm}}$. In the Standard Model alignment limit the $h$ particle is that detected by experiments at the LHC and a wide range of measurements suggest that this limit should be close to being the case~\cite{Celis:2013ixa,Chiang:2013ixa,Chen:2013rba,Chowdhury2017,Haller2018}. If there is a global $U(1)_{\rm PQ}$ symmetry then $M_A=0$ and if this is extended to $SO(3)_{\rm HF}$ then in addition one has $M_H=0$.

Topological defect solutions~\cite{Shellard1994,M&Sbook} associated with these symmetries were studied in detail in the context of the 2HDM in \cite{Brawn2011}. In particular there can be domain wall~\cite{Brawn2011, Viatic2020,Viatic2020b,Chen:2020soj,Law:2021ing}, global vortex~\cite{Brawn2011} and global monopole solutions~\cite{BATTYE2023138091}. Motivating the present study is the observation, based on field theory simulations from random initial conditions, that the vacuum is not neutral in the core of these defects~\cite{Viatic2020,Law:2021ing,BATTYE2023138091} contrary to the assumption of \cite{Brawn2011}. We will see that this has profound implications. One should note that there are also several papers discussing non-topological configurations \cite{PhysRevLett.76.356,PhysRevLett.82.2443,BACHAS1996237,PhysRevD.48.5818} and other compound structures of topological defects \cite{Eto:2020hjb,Eto:2021dca} (without any neutral vacuum violation) in the 2HDM that can be dynamically stable in certain regions of the parameter space. Here, we will focus purely only on topologically stabilised objects.

Accidental symmetries can be gauged using a covariant kinetic term
\begin{eqnarray}   D_\mu\Phi&=&\bigg[(\sigma^{0}\otimes\sigma^0)\partial_\mu+\half ig(\sigma^0\otimes\sigma^a)W_\mu^a\\\nonumber &+&\half ig^{\prime}(\sigma^0\otimes\sigma^0)Y_\mu+\half ig^{\prime\prime}(\sigma^a\otimes\sigma^0)V_\mu^a\bigg]\Phi\,,\\\nonumber
\end{eqnarray}
where $\sigma^\mu=(\sigma^0,\sigma^a)$ are the Pauli matrices including the identity. $W_{\mu}^a$ and $Y_\mu$ are the Standard Model gauge fields, with coupling constants $g$ and $g^{\prime}$ respectively, and $V_\mu^a$ are the new gauge fields associated with the accidental symmetries with coupling constant $g^{\prime\prime}$ - see \cite{supplementary} for more details on the symmetry transformations. For the purposes of this work, we set $g^\prime=0$ in order to simplify the defect solutions, but note that we do not expect any changes in the qualitative features for non-zero $g^\prime$.

Gauging the symmetries provides a natural mechanism for removing the Goldstone modes associated  with the accidental symmetries allowing for a potentially viable model. In particular in the case of a $U(1)_{\rm PQ}$ symmetry the Goldstone mode with $M_A=0$ becomes a massive gauge boson. There can be interesting models constructed with these symmetries, for example, models that can generate masses for neutrinos \cite{KO2012202,Camargo2019,GRIMUS2009117}.

In this paper we will show that the Hopf fibration associated with the Nambu monopole is realised on energetic grounds within the 2HDM when there is either a $SO(3)_{\rm HF}$ or $U(1)_{\rm PQ}$ symmetry; something that we term ``Spontaneous Hopf Fibration" (SHF). We will find monopole and vortex solutions for the case where the symmetries are gauged (although these can be easily adapted to the global limit). 

{\it Parameterizations and topology:} The 8 fields of the 2HDM can be reparameterized as 
\begin{equation}
    \Phi=\frac{v_{\rm SM}}{\sqrt{2}}e^{\half i\chi}(\sigma^0\otimes U_{L})\begin{pmatrix}
        0 & f_1 & f_+ & f_2 e^{i\xi}
    \end{pmatrix}^T\,,
    \label{ansatz}
\end{equation}
using 5 fields $f_{1,2,+}$, $\xi$ and $\chi$, and $U_{L}\in SU(2)_{\rm L}$ which has 3 degrees of freedom. The constant $v_{\rm SM}=246\,{\rm GeV}$ is the Standard Model vacuum expectation value and $f_+=0$ corresponds to a neutral vacuum with zero photon mass. These degrees of freedom can also be encoded using bilinear forms, $R^{\mu}=\Phi^{\dag}(\sigma^\mu\otimes\sigma^0)\Phi$, $n^a=-\Phi^\dag(\sigma^0\otimes\sigma^a)\Phi$ and ${\tilde R}=2\Phi_2^Ti\sigma^2\Phi_1$, which are useful for understanding the topology of the vacuum manifold and the associated defects~\cite{Ivanov2007,Brawn2011,BATTYE2023138091}. In particular, the neutral vacuum violation discovered in the core of defects in \cite{Viatic2020,BATTYE2023138091} can be traced by $R_+ = R_\mu R^\mu$, with $R_+=0$ corresponding to a neutral vacuum. One finds that two of the $U_{\rm L}$ degrees of freedom are encoded in $\hat{n}^a$, with the other associated with rotations about this axis, and the hypercharge degree of freedom, $U_{\rm Y}$, is encoded in ${\tilde R}\propto \exp[i\chi]$. In contrast, $R^{\mu}$ is invariant under the Standard Model symmetries and contains the degrees of freedom that will, in general, change the potential. The $R^0$ component is $|\Phi|^2$ and so does not contribute to the topology of the vacuum manifold. The topological non-triviality of the vacuum manifold can, therefore, be most easily extracted by looking at the remaining three-vector $R^a$, which will contain all of the degrees of freedom associated with any additional symmetry transformations, $U_H$.

A notable difference between this topology and that of the 't Hooft-Polyakov monopole~\cite{tHooft:1974kcl,Polyakov:1974ek} and Nielsen-Olesen vortex~\cite{Nielsen:1973cs} is that the topology lives in a space associated with these bilinear forms, rather than the fields themselves, which means that a half twist in field space can be topologically non-trivial and, in general, there is a factor of $2$ difference between the topological degree of a field configuration and what one might naively expect.  A consequence of this is that simple field configurations with unit winding often have discontinuities which must be resolved by the attachment of another soliton. It is for this reason that the Nambu monopole \cite{NAMBU1977505} has a string ``emanating" from one of its poles.

In the $\text{SO(3)}_{\text{HF}}$ case $R^a$ only contributes to the potential with a term $\propto R^a R^a$, so rotations between the three components of $R^a$ are a symmetry of the potential, generating an $S^2$ component of the vacuum manifold. Note that the topology is not $S^3$ as one might expect because $U_H$ cannot perform rotations about $R^a$ --- this degree of freedom is already contained within $U_L$ for rotations about $n^a$. Similarly, in the $\text{U(1)}_{\text{PQ}}$ case, the $R^3$ component splits off from the other two but there remains a symmetry for rotations between $R^1$ and $R^2$, which is responsible for an $S^1$ direction. In general, there is an additional $S^1\times S^3$ associated with the hypercharge and isospin symmetries because the degeneracy between $U_Y$ and one of the directions in $U_L$ is broken by $f_+$. Since this results in a massive photon, we avoid this scenario and choose the parameters so that $f_+=0$ in the vacuum, restoring the degeneracy so that the SM symmetries only contribute $S^3$. Therefore, the topology of the vacuum manifolds associated with the accidental symmetries are ${\cal M}=S^2\times S^3$ for the case of $SO(3)_{\rm HF}$ and ${\cal M}=S^1\times S^3$ for $U(1)_{\rm PQ}$~\cite{Brawn2011}, which admit monopoles and vortices due to the non-trivial homotopy groups $\pi_2(S^2\times S^3) = \mathbb{Z}$ and $\pi_1(S^1 \times S^3) = \mathbb{Z}$, respectively.

{\it Monopole solutions:} In the $\text{SO(3)}_{\text{HF}}$ symmetric model, using 3D spherical polar coordinates $(r,\theta,\phi)$, we can construct the monopole ansatz for the scalar field~\cite{BATTYE2023138091,supplementary}
\begin{equation}
    \Phi(r,\theta,\phi) = \frac{v_{\text{SM}}}{2\sqrt{2}}\begin{pmatrix}
        -(k+k_+)\sin\theta e^{-i\phi} \\
        (k-k_+) + (k+k_+)\cos\theta \\
        -(k-k_+) + (k+k_+)\cos\theta \\
        (k+k_+)\sin\theta e^{i\phi}
        
    \end{pmatrix} \,,
\end{equation}
where $k=k(r)$ and $k_+ = k_+(r)$ are functions constructed from $f_{1,2,+}$ that retain the ability to change the potential energy while the other degree of freedom (as well as $\xi$) is used to wind around the vacuum manifold. This ansatz has the property that $R^a = n^a = (k^2-k_+^2)\hat{r}^a$. The feature $\hat{R}^a = \hat{r}^a$ is necessary for a monopole configuration with unit winding (or related to this by a homomorphism) and this structure, by itself, would give rise to a 2HDM equivalent of the Nambu monopole --- with a divergence in the gradient energy that necessitates the emergence of a string from one of the poles. However, the isospin degrees of freedom, contained within $U_L$, can resolve this divergence when we also have that $\hat{n}^a = \hat{r}^a$, which gives the appearance of an underlying topological complexity in the structure of $n^a$, but it occurs for energetic reasons that are only indirectly topological. The degree of freedom associated with rotations about $n^a$ contains no structure --- only the possibility for global transformations --- and it is this effect that we have termed SHF. The $S^3$ of the SM becomes $S^2 \times S^1$ with the $S^2$ part inheriting the same twists as the $S^2$ of the $\text{SO(3)}_{\text{HF}}$ symmetry and the $S^1$ part containing nothing except possible global rotations. 

For the gauge fields, we choose to work in the temporal gauge such that the time components of the gauge fields are zero and make the ansatz $gW_i^a = -\frac{1}{r}h(r){\epsilon^a}_{ ij}\hat{r}^j$ and $g''V_i^a = -\frac{1}{r}H(r){\epsilon^a}_{ ij}\hat{r}^j$. After the standard rescaling to reduce the number of significant parameters in the model by two, we find that the energy functional under this ansatz is 
\begingroup
\allowdisplaybreaks
\begin{align}
    E &= \frac{4\pi v_{\text{SM}}}{g}\int r^2dr \bigg\{ \frac{1}{2}\bigg(\frac{dk}{dr}\bigg)^2 + \frac{1}{2}\bigg(\frac{dk_+}{dr}\bigg)^2 \nonumber \\
    &+ \frac{1}{8r^2}(k+k_+)^2(h+H-2)^2 + \frac{1}{8r^2}(k-k_+)^2(h-H)^2 \nonumber \\
    &+ \frac{1}{2r^2}\bigg[2\bigg(\frac{dh}{dr}\bigg)^2 + \frac{1}{r^2}h^2(2-h)^2 \bigg] + \frac{1}{2\Tilde{g}^2r^2}\bigg[2\bigg(\frac{dH}{dr}\bigg)^2 \nonumber \\
    &+ \frac{1}{r^2}H^2(2-H)^2 \bigg] + \frac{\Tilde{\lambda}}{4}\bigg[(k^2+k_+^2-1)^2 - \zeta_4 k^2k_+^2 \bigg] \bigg\} \,,
\end{align}
\endgroup
where the remaining parameters are $\Tilde{g} = g''/g$, $\Tilde{\lambda} = \lambda/g^2$ and $\zeta_4 = \lambda_4/\lambda$. Neutral vacuum violation occurs when both $k$ and $k_+$ are simultaneously non-zero and we choose, by convention, $k_+$ to be zero in the vacuum. We can perform a simple analysis (neglecting gradient energy contributions) to predict when there will be neutral vacuum violation in the core of the monopole by looking at the effective mass $m_{k_+}^2 = \lambda[2(k^2-1) - \zeta_4k^2]/4$. If a monopole were to have $k_+=0$ everywhere, then the effective mass at the core of the monopole (where $k=0$) would be $-\frac{\Tilde{\lambda}}{2}$, which is always negative and independent of $\zeta_4$. The presence of this negative mass term indicates that the energy would be reduced if $k_+\neq0$ and therefore we expect neutral vacuum violation to be a generic effect in the core of 2HDM monopoles.

\begin{figure}
    \includegraphics[width = \linewidth,trim={0.2cm 0 0cm 1.3cm},clip]{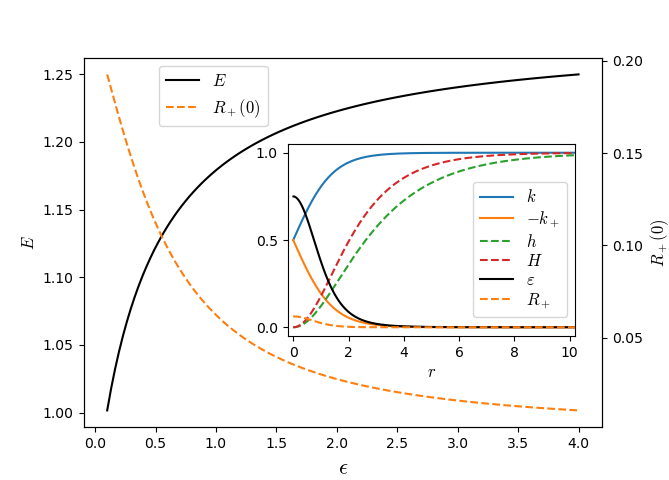}
    \caption{The variation of $E$ (black) and $R_+$ at the monopole core (dotted orange) as a function of the mass ratio $\epsilon$, with the other parameters fixed to $\Tilde{\lambda}=1$ and $\Tilde{g}^2 = 2.$ The inset plot presents the field profiles of an example solution ($\epsilon=1$) and also displays the energy density, $\varepsilon$, and $R_+$.}
    \label{fig: monopole plot}
\end{figure}

In figure \ref{fig: monopole plot} we present the energy and $R_+(0)$ (fixing $v_{\text{SM}}=1)$ as a function of the mass ratio $\epsilon = M_{H\pm}/M_h = \frac{1}{2}\sqrt{-\zeta_4}$ and we also show an example solution in the inset plot. As expected, for all values of $\epsilon$ presented here, there is neutral vacuum violation in the core of the monopole, although it decreases as $\epsilon$ grows and appears to be approaching zero, while conversely, the energy grows with $\epsilon$ and appears to be approaching a maximum. Perhaps the most noticeable feature of the solution is that $k+k_+=0$ at the centre --- in fact this is enforced by the gradient energy and is true for all values of $\epsilon$. The effects of $\Tilde{\lambda}$ can be broadly described as changing the length scale ratio between the scalar fields and the vector fields and, similarly, $\Tilde{g}$ changes the length scale ratio between the two gauge fields.

{\it Vortex solutions:} In the $\text{U(1)}_{\text{PQ}}$ symmetric case, using plane polar coordinates $(r,\theta)$, we can make the vortex ansatz $\Phi=\Phi(r,\theta)$ from equation (\ref{ansatz}) with $f_i=f_i(r)$ for $i=1,2,+$ and $\xi=\theta$ \cite{supplementary} which has a similar property to the monopole in that $\hat{R}^b = \hat{n}^b = \hat{r}^b$, where $b\in[1,2]$, although now, unlike the monopole case, $n^b=0$ in the vacuum. The remaining components are $R^3 = f_1^2-f_+^2-f_2^2$ and $n^3 = f_1^2-f_+^2+f_2^2$. We note that, in a similar way to the Nambu monopole solution, there is a string-like configuration, characterised by $\pi_1(S^2\times S^1)=Z$ and with structure only in $R^b$, where the divergence in the gradient energy is resolved by attaching a domain wall to one side. This is related to an unstable (due to the tension in the wall) configuration in the SM where $\hat{n}^b = \hat{r}^b$. Once again, for the 2HDM configuration, we can use the isospin rotations to resolve the divergence without the domain wall. The SHF acts here, again, to split $S^3 \to S^2 \times S^1$, but now it is the $S^1$ part that inherits the twists of the $\text{U(1)}_{\text{PQ}}$. 

Now we choose to make a gauge transformation that absorbs the phase winding of the scalar field into $W_i^a$ and $V_i^3$ (the only new gauge field in the $\text{U(1)}_{\text{PQ}}$ model) so that we can make the ansatz $gW_i^a = \frac{1}{r}[h_1(r)\hat{x}^a + (1-h_3(r))\hat{z}^a]\hat{\theta}_i + h_2(r)\hat{y}^a\hat{r}_i$ and $g''V_i^3 = \frac{1}{r}(1-H(r))\hat{\theta}_i$. Again, if we make the standard rescalings then we can express the energy per unit length of the string as 
\begin{align}
    E &= 2\pi v_{\text{SM}}^2\eta_1^2\int rdr\bigg\{ \frac{1}{2}\bigg(\frac{df_1}{dr}\bigg)^2 + \frac{1}{2}\bigg(\frac{df_+}{dr}\bigg)^2 + \frac{1}{2}\bigg(\frac{df_2}{dr}\bigg)^2 \nonumber \\
    &- \frac{1}{2}h_2\bigg(f_+\frac{df_2}{dr} - f_2\frac{df_+}{dr}\bigg) + \frac{1}{8}\bigg(\frac{h_1^2}{r^2}+ h_2^2 \nonumber \\
    &+ \frac{1}{r^2}(h_3-H)^2\bigg)(f_1^2+f_+^2) - \frac{1}{2r^2}h_1(1-H)f_+f_2 \nonumber \\
    &+ \frac{1}{8}\bigg(\frac{h_1^2}{r^2} + h_2^2 + \frac{1}{r^2}(2-h_3-H)^2\bigg)f_2^2 + \frac{1}{2r^2}\bigg[ \bigg(\frac{dh_1}{dr}\bigg)^2 \nonumber \\
    &+ \bigg(\frac{dh_3}{dr}\bigg)^2 - 2h_2\bigg(h_1\frac{dh_3}{dr}+(1-h_3)\frac{dh_1}{dr}\bigg) + h_1^2h_2^2 \nonumber \\
    &+ (1-h_3)^2h_2^2 \bigg] + \frac{1}{2\Tilde{g}^2r^2}\bigg(\frac{dH}{dr}\bigg)^2 + \frac{\Tilde{\lambda}_1}{4}\bigg[(f_1^2-1)^2 \nonumber \\
    &+ \zeta_2(f_+^2+f_2^2-\Tilde{\eta}^2)^2 + \zeta_3f_1^2(f_+^2+f_2^2) + \zeta_4f_1^2f_2^2\bigg] \bigg\},
\end{align}
%
%
where $\Tilde{g} = g^{\prime\prime}/g$, $\Tilde{\eta} = \eta_2/\eta_1$, $\Tilde{\lambda}_1 = \lambda_1/g^2$ and $\zeta_i = \lambda_i/\lambda_1$ so that the model is left with 6 parameters ($\eta_i$ is defined by the relationship $\mu_i^2 = \lambda_i\eta_i^2v_{\text{SM}}^2$). We can perform an effective mass analysis here too, just as in the monopole case. The relevant effective mass is $m_{f_+}^2 ={\Tilde\lambda}_1[2\zeta_2(f_2^2-\Tilde{\eta}^2) + \zeta_3f_1^2]/4$ and if the string were to have $f_+=0$ everywhere, as well as $f_1(0)=1$ and $f_2(0)=0$, then this would take the value $\Tilde{\lambda}_1[-2\zeta_2\Tilde{\eta}^2 + \zeta_3]/4$, so the neutral vacuum violation in the core is not generic but depends upon the sign of $\zeta_3-2\zeta_2\Tilde{\eta}^2$. Note that, $f_2(0)=0$ is enforced by the winding of the string but $f_1(0)=1$ is an overly simplistic assumption, even if $f_+=0$ everywhere, so this approach will be less accurate for vortices than for the monopoles.

\begin{figure}[t]
    \includegraphics[width = \linewidth,trim={0.5cm 0.2cm 0cm 1.3cm},clip]{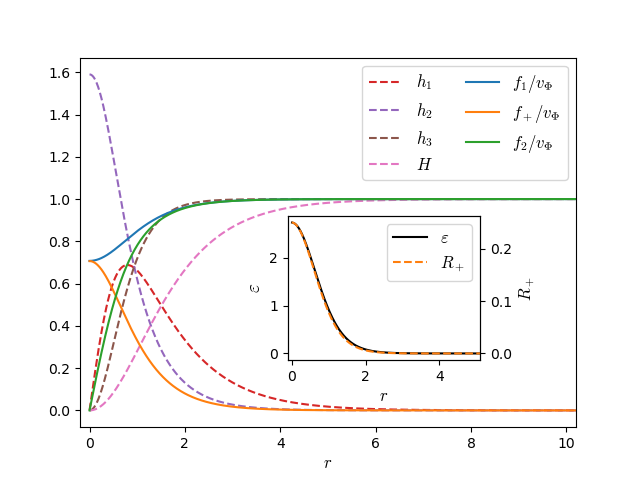}
    \caption{The field profiles for a string solution with $\Tilde{\lambda}_1 = 1$, $\Tilde{g}^2=1$, $\Tilde{\eta}^2 = 1$, $\zeta_2 = 1$, $\zeta_3 = -0.4$ and $\zeta_4=-0.8$ ---  the last four parameter choices corresponding to the alignment limit with $\epsilon=1$, $\delta=2$, $\tan\beta=1$. The inset plot displays the energy density, $\varepsilon$, and $R_+$ for the same solution.}
    \label{fig: string profiles}
\end{figure}

\begin{figure}[t]
    \includegraphics[width = \linewidth,trim={0.5cm 0cm 0cm 1.3cm},clip]{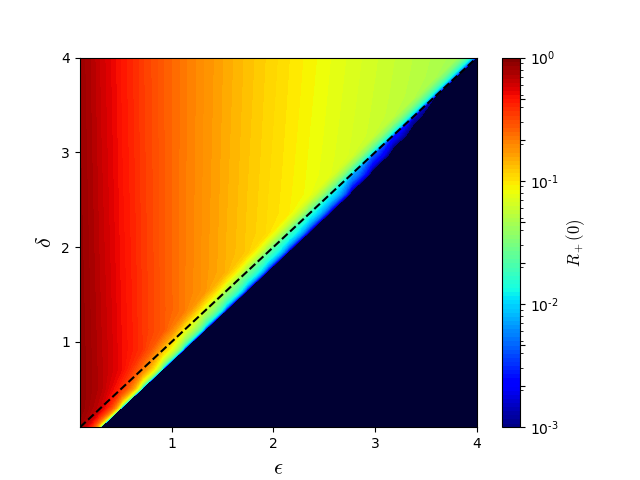}
    \caption{A contour plot showing how $R_+$ at the string core varies with the mass ratios $\epsilon$ and $\delta$ in the alignment limit and with $\Tilde{\lambda_1}=\Tilde{g}^2=\tan\beta=1$. Note that the dark blue colour on most of the lower right region of the plot is off the bottom of the colour scale - corresponding to a neutral vacuum.}
    \label{fig: string contour}
\end{figure}

In figure \ref{fig: string profiles} we present the field profiles of a string solution with a separate inset plot showing the energy density and $R_+$ for the same solution and in figure \ref{fig: string contour} we show how $R_+$, at the core of the string, varies with the mass ratios $\epsilon = M_{H\pm}/M_h$ and $\delta = M_H/M_h$, in the alignment limit and with $\tan\beta=1$ (which sets the vacuum values of the fields so that $|\Phi_1|^2=|\Phi_2|^2 = v_\Phi^2$). Note that we have rescaled $R_+ \to R_+/(v_\Phi v_{\text{SM}})^4$ implicitly in these plots.

From the profiles we see that the solution has retained a feature that is similar to one observed for the monopole solutions, namely that $f_1=f_+$ at the centre. However, this is no longer guaranteed by the gradient energy and, in fact, is only an approximate equality that occurs in a subset of the parameter space. 
\noindent The parameters $\Tilde{\lambda}_1$ and $\Tilde{g}^2$ play a very similar role to the equivalent parameters from the monopole case but the other parameters have more complicated effects on the solutions that are difficult to broadly summarise in this way. 
\noindent From the contour plot we can see that there is a clear transition across the line which is approximately $\epsilon=\delta$, corresponding to $\zeta_3=2$. Due to our fixed parameter choices we have $\zeta_2=\Tilde{\eta}^2=1$  and therefore this is consistent with the behaviour that we predicted from the effective mass analysis.

{\it Discussion and conclusions:} The solutions that we have presented in this paper are evidence of a new mechanism --- that we have called ``Spontaneous Hopf Fibration" --- at work in the 2HDM. It allows for a topologically non-trivial subspace of the vacuum manifold to imprint itself onto another, topologically trivial section of the manifold. This results in solitons that appear to have topological structure in the SM degrees of freedom but, in fact, it is purely caused by energetics. The coupling between the $S^3$ of the SM to the rest of the vacuum manifold, in the gradient energy term, causes a Hopf fibration of the space $S^3 \to S^2 \times S^1$, with the appropriate component of this space taking on structure to match the winding around $S^2$ for monopoles and $S^1$ for strings.

In \cite{BATTYE2023138091} we present evidence from simulations of the global 2HDM model in which stable monopoles form that have neutral vacuum violation in the cores and a structure in the bilinear vectors that is the same as what one would expect from the solutions presented here. In \cite{supplementary} we do the same for the case of strings. These simulations suggest that the solutions we have found are those most relevant to the study of topological defects in the 2HDM.

A phenomenologically relevant consequence of the SHF in the 2HDM is that it allows for the neutral vacuum condition to be violated inside the core of the defects, generating a non-zero mass for the photon. In the case of strings, this is dependent upon the choice of parameters, however in the monopole case it is predicted to always occur if the gradient energy contributions are neglected. The interaction between photons and superconducting defects has been analysed in \cite{Battye:2021dyq} for a toy model but this work has opened up the possibility for novel interactions between standard model particles and 2HDM defects which is deserving of more investigation. In \cite{BATTYE2023138091} it was observed that the additional structure of global 2HDM monopoles did not affect the scaling of their number density, but other potential cosmological consequences warrant further studies of these defects.

We would like to conclude by emphasising that, although we have discussed SHF in the context of the 2HDM, we suggest that it could be a more general effect that can occur in other models that have vacuum manifolds constructed from coupled subspaces, when at least one of them is topologically non-trivial.

{\it Acknowledgements:} We would like to thank Jeff Forshaw for helpful comments on the phenomenological viability of these models. RB would like to thank Apostolos Pilaftsis, Gary Brawn and Dominic Viatic for their collaboration on the 2HDM which ultimately lead to this work.

\bibliographystyle{apsrev4-1}
\bibliography{refs.bib}

\end{document}


\title{Spontaneous Hopf fibration in the two-Higgs-doublet model : Supplemental material}
 
\author{R.~A.~Battye}
\email[]{richard.battye@manchester.ac.uk}
\affiliation{%
Jodrell Bank Centre for Astrophysics, Department of Physics and Astronomy, University of Manchester, Manchester, M13 9PL, U.K.
}

\author{S.~J.~Cotterill}
\email[]{steven.cotterill@manchester.ac.uk}
\affiliation{%
Jodrell Bank Centre for Astrophysics, Department of Physics and Astronomy, University of Manchester, Manchester, M13 9PL, U.K.
}

\label{firstpage}

\date{\today}

\maketitle

\section{The Hopf Fibration}

To understand the Hopf fibration it is useful to first understand the Hopf map, which is a many-to-one map from the 3-sphere onto the 2-sphere: $S^3\to S^2$. One can embed the 3-sphere in $\mathbb{C}^2$ using the \textit{Hopf coordinates}
\begin{equation}
    z = \begin{pmatrix}
        z_0 \\
        z_1
    \end{pmatrix} = \begin{pmatrix}
        \cos\eta e^{i\xi_0} \\
        \sin\eta e^{i\xi_1}
    \end{pmatrix} \,,
\end{equation}
where $0\leq\eta\leq\pi/2$ and $0\leq\xi_0,\xi_1\leq2\pi$ with $z^\dagger z=|z_0|^2+|z_1|^2=1$. The Hopf map can then be expressed as
\begin{equation}
    n^a = z^\dagger\sigma^a z = \begin{pmatrix}
        \sin2\eta\cos(\xi_1-\xi_0) \\
        \sin2\eta\sin(\xi_1-\xi_0) \\
        \cos2\eta
    \end{pmatrix}\,.
\end{equation}
We can now define new angular coordinates $\alpha = 2\eta$, $\beta = \xi_1-\xi_0$ and $\zeta = \xi_0 + \xi_1$, with $0\leq\alpha\leq\pi$, $0\leq\beta\leq 2\pi$ and $0\leq\zeta\leq4\pi$, where we have used the periodicity of $\xi_0$ and $\xi_1$ to set the ranges of $\beta$ and $\zeta$. In these new coordinates it becomes clear that $n^a$ describes a two-sphere that does not depend upon $\zeta$, which itself describes different points on $S^3$ (great circles) that all get mapped to the same point on $S^2$.

\begin{figure}[!t]
   \centering
   \captionsetup{justification=centering}
    \subfloat[$\alpha = \pi/5$]{
        \centering
        \includegraphics[trim={4.4cm 2cm 4cm 1cm},clip,width=0.48\linewidth]{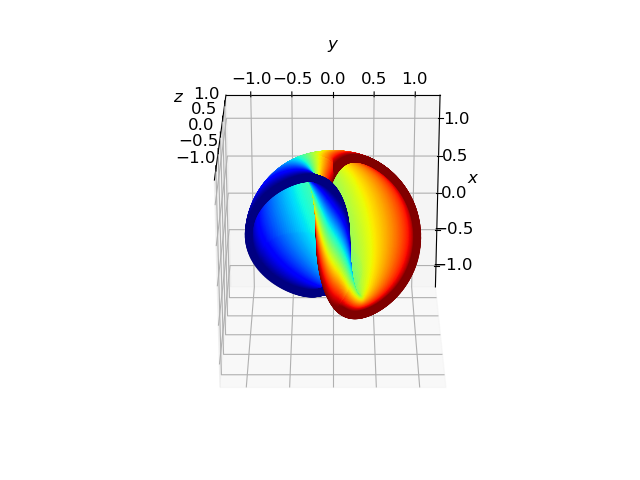}
    }
    \subfloat[$\alpha = 2\pi/5$]{
        \centering
        \includegraphics[trim={4.4cm 2cm 4cm 1cm},clip,width=0.48\linewidth]{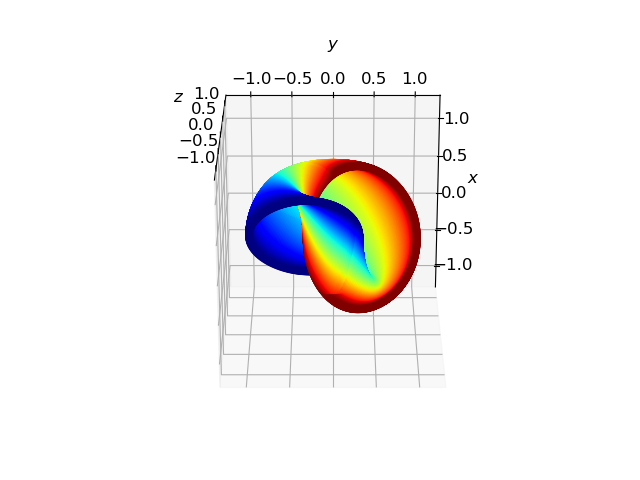}
    }\hfill
    \subfloat[$\alpha = 3\pi/5$]{
        \centering
        \includegraphics[trim={4.4cm 2cm 4cm 1cm},clip,width=0.48\linewidth]{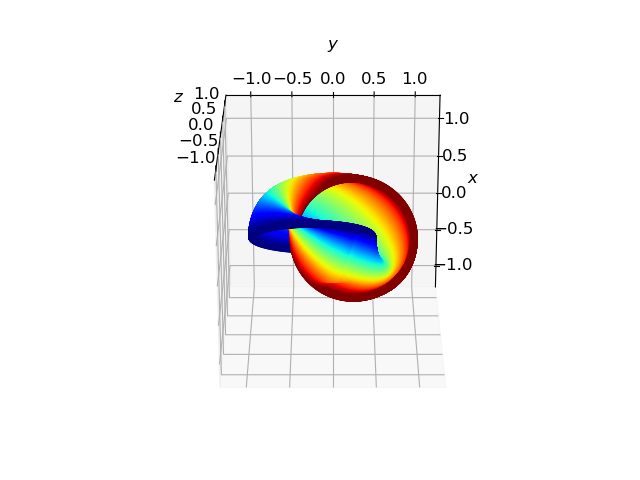}
    }
    \subfloat[$\alpha = 4\pi/5$]{
        \centering
        \includegraphics[trim={4.4cm 2cm 4cm 1cm},clip,width=0.48\linewidth]{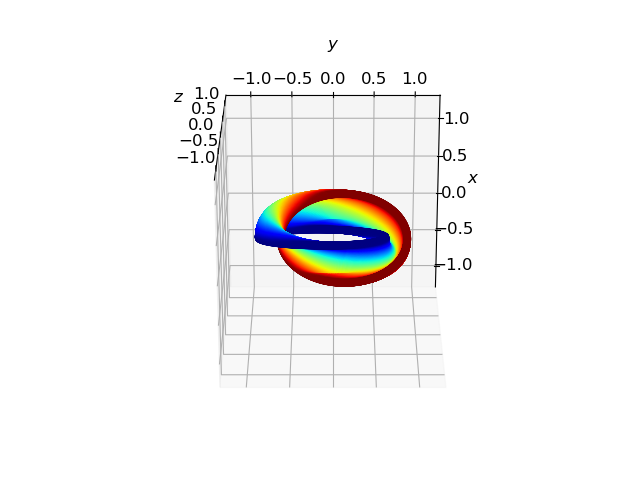}
    }\hfill
    \captionsetup{justification=Justified}
    \caption{A visualisation of the Hopf fibration for four different values of $\alpha$, representing half the range of $\beta$ with colours from $\beta=\pi$ (blue) to $\beta=2\pi$ (red) and the full range of $\zeta$ is displayed. The angles $\alpha$ and $\beta$ represent unique positions in $n^a$ (on $S^2$) while $\zeta$ parameterises the $S^1$ direction, which is clearly visible in these plots as lines of constant colour.}
    \label{fig: Hopf}
\end{figure}

The Hopf fibration is therefore an example of a fibre bundle where $S^3 \cong S^2 \times S^1$ locally, but not globally, and the $S^2$ corresponds to unique points on $n^a$, while $S^1$ is represented by the remaining angular coordinate $\zeta$. This can be visualised by performing a stereographic projection from the point $z_0=1$, $z_1=0$ onto $\mathbb{R}^3$ and compactifying the full, infinite space into a sphere of radius 1 by re-defining the spherical radius as $r^\prime = \tanh(r)$. We plot the resulting surfaces for four different values of $\alpha$ in Figure \ref{fig: Hopf} and represent the value of $\beta$ with colour so that the $S^1$ directions are represented by lines of constant colour.

\section{Symmetries of the model}

With the inclusion of accidental symmetries, the field can be parameterized as 
\begin{equation}
\Phi=\frac{v_{\rm SM}}{\sqrt{2}}e^{\half i\chi}(U_{\rm H}\otimes U_{\rm L})\Bar{\Phi}\,,
\end{equation}
in such a way that all of the degrees of freedom are separated into those that do, ${\bar\Phi}$, and do not, $\chi$, $U_{\rm L}$ and $U_{\rm H}$, change the potential energy. For the case of an $\text{SO(3)}_{\text{HF}}$ symmetry, $U_H$ can be any element of SU(2). However the full set of SU(2) matrices would represent a double-cover of the field (e.g $U_{\rm H} \to -U_{\rm H}$ can already be covered by $U_{\rm L} \to -U_{\rm L}$) so this is typically reduced down to $\text{SU(2)}/\mathbb{Z}_2\cong\text{SO(3)}$. Any field configuration can be generated from symmetry transformations applied to $\Bar{\Phi}= \begin{pmatrix} 0 & k & k_+ & 0\end{pmatrix}^T$. For the $\text{U(1)}_{\text{PQ}}$ case, the symmetry is generated by matrices of the form $U_H \in \text{diag}\begin{pmatrix}e^{\half i\gamma} & e^{-\half i\gamma}\end{pmatrix}$ for $0<\gamma<2\pi$ and $\Bar{\Phi} = \begin{pmatrix}0 & f_1 & f_+ & f_2\end{pmatrix}^T$.

The new gauge symmetry is $\Phi \to (U_H\otimes\sigma^0)\Phi$ and $V_\mu^a \to V_\mu^b\mathcal{R}^{ba} + v_i^a/g^{\prime\prime}$ where $\mathcal{R}^{ab}_H$ is a rotation matrix defined by $U_H^\dagger\sigma^a U_H = \mathcal{R}^{ab}_H\sigma^b$ and $v_i^a$ is defined by $U_H^\dagger\partial_\mu U_H = \half iv_i^a\sigma^a$. This expression can be used for both symmetry cases under consideration here but in the $\text{SO(3)}_{\text{HF}}$ case all 3 components of $V_\mu^a$ are used, with $a \in \{1,2,3\}$, whereas in the $\text{U(1)}_{\text{PQ}}$ case only $V_\mu^3$ is used with $V_1^\mu=V_2^\mu\equiv 0$. 

The effect of the new symmetry transformation on the bilinear vector, $R^a$, is simply a rotation $R^{a\prime} = \mathcal{R}^{ab}_HR^b$, which corresponds to a 3D rotation in the $\text{SO(3)}_{\text{HF}}$ symmetric case and a 2D rotation about the $3$ axis in the $\text{U(1)}_{\text{PQ}}$ case. Likewise, the $n^a$ vector rotates under isospin transformations  $n^{a\prime} = \mathcal{R}^{ab}_Ln^b$ where $U_L^\dagger\sigma^a U_L = \mathcal{R}^{ab}_L\sigma^b$.



\section{Monopole solution}

Our ansatz is generated by applying 
\begin{equation}
    U_H = U_L = \begin{pmatrix}
        \cos\half\theta & -\sin\half\theta e^{-i\phi} \\
        \sin\half\theta e^{i\phi} & \cos\half\theta
    \end{pmatrix}\,,
\end{equation}
to $\Bar{\Phi}$, where $k$ and $k_+$ are now functions of the radius only. For the gauge fields, we use an expression which will cancel the gradient energy of the monopole far from its core and that leaves the system with a spherically symmetric energy density, and therefore functions in the static equations of motion that only depend on the radius. Using this ansatz, the energy of the configuration is found to be given by
\begin{align}
    E &= 4\pi\int r^2dr \bigg\{ \frac{1}{2}v_{\text{SM}}^2\bigg[ \bigg(\frac{dk}{dr}\bigg)^2 + \bigg(\frac{dk_+}{dr}\bigg)^2 \nonumber \\
    &+ \frac{1}{4r^2}(k+k_+)^2(h+H-2)^2 + \frac{1}{4r^2}(k-k_+)^2(h-H)^2 \bigg] \nonumber \\
    &+ \frac{1}{2g^2r^2}\bigg[2\bigg(\frac{dh}{dr}\bigg)^2 + \frac{1}{r^2}h^2(2-h)^2 \bigg] + \frac{1}{2g''^2r^2}\bigg[2\bigg(\frac{dH}{dr}\bigg)^2 \nonumber \\
    &+ \frac{1}{r^2}H^2(2-H)^2 \bigg] + \frac{1}{4}v_{\text{SM}}^4\bigg[ \lambda(k^2+k_+^2-1)^2 \nonumber \\
    &- \lambda_4 k^2k_+^2 \bigg] \bigg\} \,,
\end{align}
where $\mu^2 = \mu_1^2 = \mu_2^2$, $\lambda = \lambda_1 = \lambda_2$ and $v_{\text{SM}}^2 = \mu^2/\lambda$. We can remove two of the parameters by rescaling lengths with $\Tilde{r} = gv_{\text{SM}} r$ which gives
\begingroup
\allowdisplaybreaks
\begin{align}
    E &= \frac{4\pi v_{\text{SM}}}{g}\int \Tilde{r}^2d\Tilde{r} \bigg\{ \frac{1}{2}\bigg[ \bigg(\frac{dk}{d\Tilde{r}}\bigg)^2 + \bigg(\frac{dk_+}{d\Tilde{r}}\bigg)^2 \nonumber \\
    &+ \frac{1}{4\Tilde{r}^2}(k+k_+)^2(h+H-2)^2 + \frac{1}{4\Tilde{r}^2}(k-k_+)^2(h-H)^2 \bigg] \nonumber \\
    &+ \frac{1}{2\Tilde{r}^2}\bigg[2\bigg(\frac{dh}{d\Tilde{r}}\bigg)^2 + \frac{1}{\Tilde{r}^2}h^2(2-h)^2 \bigg] + \frac{1}{2\Tilde{g}^2\Tilde{r}^2}\bigg[2\bigg(\frac{dH}{d\Tilde{r}}\bigg)^2 \nonumber \\
    &+ \frac{1}{\Tilde{r}^2}H^2(2-H)^2 \bigg] + \frac{\Tilde{\lambda}}{4}\bigg[(k^2+k_+^2-1)^2 \nonumber \\
    &- \zeta_4 k^2k_+^2 \bigg] \bigg\} \,,
\end{align}
\endgroup
where the remaining parameters are $\Tilde{g} = g''/g$, $\Tilde{\lambda} = \lambda/g^2$ and $\zeta_4 = \lambda_4/\lambda$. The resulting set of static equations of motion for the gauged monopole are
\begingroup
\allowdisplaybreaks
\begin{align}
    &\frac{d^2k}{d\Tilde{r}^2} + \frac{2}{\Tilde{r}}\frac{dk}{d\Tilde{r}} - \frac{1}{4\Tilde{r}^2}\Big[ (k+k_+)(h+H-2)^2 \nonumber \\
    &+ (k-k_+)(h-H)^2 \Big] - \Tilde{\lambda}\bigg[(k^2+k_+^2-1)k \nonumber \\
    &- \frac{1}{2}\zeta_4kk_+^2\bigg] = 0 \,, \\
    &\frac{d^2k_+}{d\Tilde{r}^2} + \frac{2}{\Tilde{r}}\frac{dk_+}{d\Tilde{r}} - \frac{1}{4\Tilde{r}^2}\Big[ (k+k_+)(h+H-2)^2 \nonumber \\
    &- (k-k_+)(h-H)^2 \Big] - \Tilde{\lambda}\bigg[(k^2+k_+^2-1)k_+ \nonumber \\
    &- \frac{1}{2}\zeta_4k^2k_+\bigg] = 0 \,, \\
    &\frac{d^2h}{d\Tilde{r}^2} - \frac{1}{\Tilde{r}^2}h(1-h)(2-h)- \frac{1}{8}\Big[ (k+k_+)^2(h+H-2) \nonumber \\
    &+ (k-k_+)^2(h-H) \Big] = 0 \,, \\
    &\frac{d^2H}{d\Tilde{r}^2} - \frac{1}{\Tilde{r}^2}H(1-H)(2-H) - \frac{\Tilde{g}^2}{8}\Big[ (k+k_+)^2(h+H-2) \nonumber \\
    &- (k-k_+)^2(h-H) \Big] = 0 \,.
\end{align}
\endgroup
The boundary conditions for this system of equations are $k(0)=-k_+(0)$, $h(0)=H(0)$, $k_+(\infty)=0$ and $k(\infty)=h(\infty)=H(\infty)=1$ where we have assumed that $\zeta_4<0$, so that the minimum energy vacuum state is one where $f_+=0$, to prevent a massive photon.

\section{Vortex solution}

Similar to the monopole case, our ansatz for the scalar field can be generated by the application of $U_H \otimes U_L$ onto $\Bar{\Phi} = \begin{pmatrix}0 & f_1 & f_+ & f_2\end{pmatrix}^T$ where
\begin{equation}
    U_H = U_L = \begin{pmatrix}
        e^{-\half i\theta} & 0 \\
        0 & e^{\half i \theta}
    \end{pmatrix}\,.
\end{equation}
However, the gradient energy of this string ansatz is much more complicated than in the monopole case, due to the additional scalar degree of freedom. Therefore, we have found it simpler to make a gauge transformation to absorb these windings in $U_H$ and $U_L$ into $V_i^3$ and $W_i^3$ respectively. The rest of the ansatz for the gauge fields was chosen so that the energy density respects the cylindrical symmetry of the string, which gives rise to one-dimensional equations of motion, and also so that any additional gauge components which can reduce the energy of the system are not fixed to zero.
The resulting energy is
\begingroup
\allowdisplaybreaks
\begin{align}
    E &= 2\pi\int rdr \bigg\{ \frac{1}{2}v_{\text{SM}}^2\bigg[ \bigg(\frac{df_1}{dr}\bigg)^2 + \bigg(\frac{df_+}{dr}\bigg)^2 + \bigg(\frac{df_2}{dr}\bigg)^2 \nonumber \\
    &- h_2\bigg(f_+\frac{df_2}{dr} - f_2\frac{df_+}{dr}\bigg) + \frac{1}{4}\bigg(\frac{h_1^2}{r^2} + h_y^2 \nonumber \\
    &+ \frac{1}{r^2}(h_3-H)^2\bigg)(f_1^2+f_+^2) - \frac{1}{r^2}h_1(1-H)f_+f_2 \nonumber \\
    &+ \frac{1}{4}\bigg(\frac{h_1^2}{r^2} + h_2^2 + \frac{1}{r^2}(2-h_3-H)^2\bigg)f_2^2 \bigg] \nonumber \\
    &+ \frac{1}{2g^2r^2}\bigg[ \bigg(\frac{dh_1}{dr}\bigg)^2 + \bigg(\frac{dh_3}{dr}\bigg)^2 - 2h_2\bigg(h_1\frac{dh_3}{dr} \nonumber \\
    &+ (1-h_3)\frac{dh_1}{dr}\bigg) + (h_1^2+(1-h_3)^2)h_2^2 \bigg] \nonumber \\
    &+ \frac{1}{2g''^2r^2}\bigg(\frac{dH}{dr}\bigg)^2 + \frac{1}{4}v_{\text{SM}}^4\bigg[ \lambda_1(f_1^2-\eta_1^2)^2 \nonumber \\
    &+ \lambda_2(f_+^2+f_2^2-\eta_2^2)^2 + \lambda_3f_1^2(f_+^2+f_2^2) + \lambda_4f_1^2f_2^2 \bigg] \bigg\} \,,
\end{align}
\endgroup
where $\mu_1^2 = \lambda_1\eta_1^2v_{\text{SM}}^2$ and $\mu_2^2 = \lambda_2\eta_2^2v_{\text{SM}}^2$. We can now rescale lengths with $\Tilde{r} = gv_{\text{SM}}\eta_1 r$ and field magnitudes with $f_i = \eta_1 \Tilde{f}_1$ and $h_2 = gv_{\text{SM}}\eta_1\Tilde{h}_2$ to give
\begin{align}
    E &= 2\pi v_{\text{SM}}^2\eta_1^2\int \Tilde{r}d\Tilde{r}\bigg\{ \frac{1}{2}\bigg(\frac{d\Tilde{f}_1}{d\Tilde{r}}\bigg)^2 + \frac{1}{2}\bigg(\frac{d\Tilde{f}_+}{d\Tilde{r}}\bigg)^2 + \frac{1}{2}\bigg(\frac{d\Tilde{f}_2}{d\Tilde{r}}\bigg)^2 \nonumber \\
    &- \frac{1}{2}\Tilde{h}_2\bigg(\Tilde{f}_+\frac{d\Tilde{f}_2}{d\Tilde{r}} - \Tilde{f}_2\frac{d\Tilde{f}_+}{d\Tilde{r}}\bigg) + \frac{1}{8}\bigg(\frac{h_1^2}{\Tilde{r}^2}+\Tilde{h}_2^2 \nonumber \\
    &+ \frac{1}{\Tilde{r}^2}(h_3-H)^2\bigg)(\Tilde{f}_1^2+\Tilde{f}_+^2) - \frac{1}{2\Tilde{r}^2}h_1(1-H)\Tilde{f}_+\Tilde{f_2} \nonumber \\
    &+ \frac{1}{8}\bigg(\frac{h_1^2}{\Tilde{r}^2} + \Tilde{h}_2^2 + \frac{1}{\Tilde{r}^2}(2-h_3-H)^2\bigg)\Tilde{f}_2^2 + \frac{1}{2\Tilde{r}^2}\bigg[ \bigg(\frac{dh_1}{d\Tilde{r}}\bigg)^2 \nonumber \\
    &+ \bigg(\frac{dh_3}{d\Tilde{r}}\bigg)^2 - 2\Tilde{h}_2\bigg(h_1\frac{dh_3}{d\Tilde{r}}+(1-h_3)\frac{dh_1}{d\Tilde{r}}\bigg) + (h_1^2 \nonumber \\
    &+(1-h_3)^2)\Tilde{h}_2^2 \bigg] + \frac{1}{2\Tilde{g}^2\Tilde{r}^2}\bigg(\frac{dH}{d\Tilde{r}}\bigg)^2 + \frac{\Tilde{\lambda}_1}{4}\bigg[(\Tilde{f}_1^2-1)^2 \nonumber \\
    &+ \zeta_2(\Tilde{f}_+^2+\Tilde{f}_2^2-\Tilde{\eta}^2)^2 + \zeta_3\Tilde{f}_1^2(\Tilde{f}_+^2+\Tilde{f}_2^2) + \zeta_4\Tilde{f}_1^2\Tilde{f}_2^2\bigg] \bigg\} \,,
\end{align}
where $\Tilde{g} = g''/g$, $\Tilde{\eta} = \eta_2/\eta_1$, $\Tilde{\lambda}_1 = \lambda_1/g^2$ and $\zeta_i = \lambda_i/\lambda_1$ so that the model has 6 significant parameters. The equations of motion are (dropping the tilde on all quantities for ease and clarity)
\begingroup
\allowdisplaybreaks
\begin{align}
    &\frac{d^2f_1}{dr^2} + \frac{1}{r}\frac{df_1}{dr} - \frac{1}{4}\bigg(\frac{h_1^2}{r^2} + h_2^2 + \frac{1}{r^2}(h_3-H)^2\bigg)f_1 \nonumber \\
    &- \lambda_1\bigg[(f_1^2-1) + \frac{1}{2}\zeta_3(f_+^2+f_2^2) + \frac{1}{2}\zeta_4f_2^2 \bigg]f_1 = 0 \,, \\
    &\frac{d^2f_+}{dr^2} + \frac{1}{r}\frac{df_+}{dr} + h_2\frac{df_2}{dr} + \frac{1}{2}f_2\frac{dh_2}{dr} + \frac{1}{2r}h_2f_2 \nonumber \\
    &+ \frac{1}{2r^2}h_1(1-H)f_2 - \frac{1}{4}\bigg(\frac{h_1^2}{r^2} + h_2^2 +\frac{1}{r^2}(h_3-H)^2\bigg)f_+ \nonumber \\
    &- \lambda_1\bigg[\zeta_2(f_+^2+f_2^2-\eta^2) + \frac{1}{2}\zeta_3f_1^2 \bigg]f_+ = 0 \,, \\
    &\frac{d^2f_2}{dr^2} + \frac{1}{r}\frac{df_2}{dr} - h_2\frac{df_+}{dr} - \frac{1}{2}f_+\frac{dh_2}{dr} - \frac{1}{2r}h_2f_+ \nonumber \\
    &+ \frac{1}{2r^2}h_1(1-H)f_+ - \frac{1}{4}\bigg(\frac{h_1^2}{r^2} + h_2^2 \nonumber \\
    &+ \frac{1}{r^2}(2-h_3-H)^2\bigg)f_2 - \lambda_1\bigg[\zeta_2(f_+^2+f_2^2-\eta^2) \nonumber \\
    &+ \frac{1}{2}\zeta_3f_1^2 + \frac{1}{2}\zeta_4f_1^2 \bigg]f_2 = 0 \,, \\
    &\frac{d^2h_1}{dr^2} - \frac{1}{r}\frac{dh_1}{dr} + 2h_2\frac{dh_3}{dr} - (1-h_3)\frac{dh_2}{dr} + \frac{1}{r}h_2(1-h_3) \nonumber \\
    &- h_2^2h_1 - \frac{1}{4}\bigg[(f_1^2+f_+^2+f_2^2)h_1 - 2(1-H)f_+f_2\bigg] = 0 \,, \\
    &\frac{1}{r^2}\bigg[(h_1^2+(1-h_3)^2)h_2 - h_1\frac{dh_3}{dr} - (1-h_3)\frac{dh_1}{dr}\bigg] \nonumber \\
    &+ \frac{1}{4}(f_1^2+f_+^2+f_2^2)h_3 - \frac{1}{2}\bigg(f_+\frac{df_2}{dr} - f_2\frac{df_+}{dr}\bigg) = 0 \,, \\
    &\frac{d^2h_3}{dr^2} - \frac{1}{r}\frac{dh_3}{dr} - 2h_2\frac{dh_1}{dr} - h_1\frac{dh_2}{dr} + \frac{1}{r}h_1h_2 \nonumber \\
    &+ h_2^2(1-h_3) - \frac{1}{4}\bigg[(h_3-H)(f_1^2+f_+^2) \nonumber \\
    &- (2-h_3-H)f_2^2\bigg] =0 \,, \\
    &\frac{d^2H}{dr^2} - \frac{1}{r}\frac{dH}{dr} + \frac{g^2}{4}\bigg[(h_3-H)(f_1^2+f_+^2) \nonumber \\
    &+ (2-h_3-H)f_2^2 - 2h_1f_+f_2\bigg] = 0 \,.
\end{align}
\endgroup
We use the boundary conditions $f_1'(0)=f_+'(0)=f_2(0)=h_1(0)=h_3(0)=H(0)=0$ and $h_2(0) = h_1'(0)$ at the centre of the string, while $f_1^2(\infty) = \zeta_2(1-\Bar{\zeta}_{34}\eta^2)/(\zeta_2-\Bar{\zeta}^2_{34})$, $f_2^2(\infty) = (\zeta_2\eta^2-\Bar{\zeta}_{34})/(\zeta_2 - \Bar{\zeta}^2_{34})$, $f_+(\infty) = h_1(\infty) = h_2(\infty) = 0$ and $h_3(\infty) = H(\infty) = 1$, where we have defined $\Bar{\zeta}_{34} = (\zeta_3+\zeta_3)/2$ and once again assumed that $\zeta_4<0$. This corresponds to a finite energy solution with the possibility of neutral vacuum violation at the core of the string.

\section{String solution from random initial conditions}

\begin{figure}[t!]
    \centering
    \includegraphics[trim={0cm 0cm 0.5cm 1.0cm},clip,width=\linewidth]{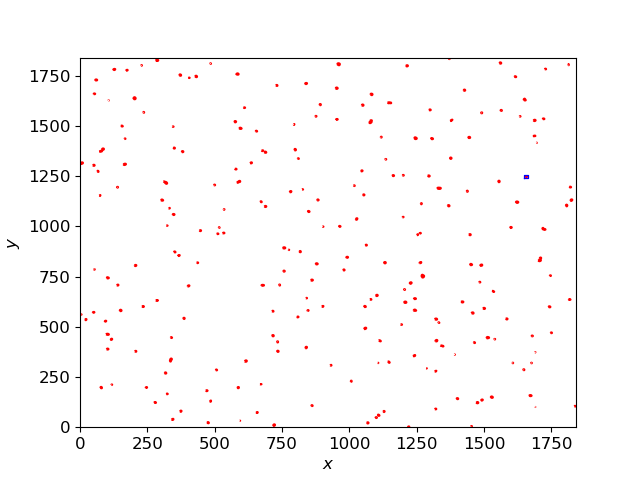}
    \caption{Surfaces with $R_+=0.1$ in a simulation with random initial conditions. There is a small blue box close to the right hand side that surrounds one of the strings. We present a close-up of this box in figure \ref{fig: bilinear vectors} that shows the structure of the bilinear vectors around the string.}
    \label{fig: Rp}
\end{figure}

\begin{figure}[H]
    \centering
    \subfloat{
        \centering
        \includegraphics[trim={0cm 0cm 0.5cm 1.0cm},clip,width=\linewidth]{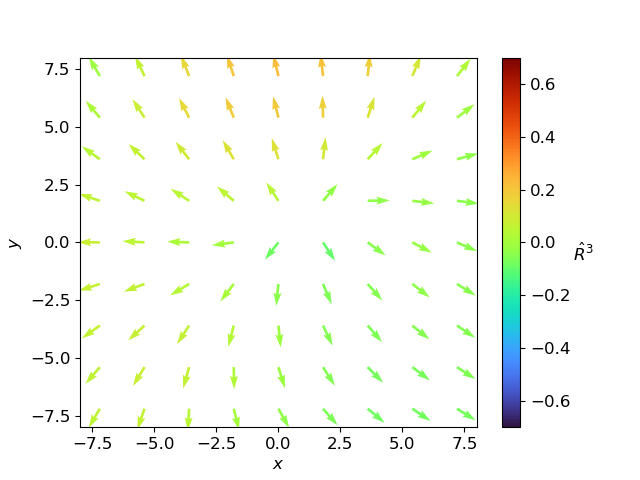}
    }\hfill
    \subfloat{
        \includegraphics[trim={0cm 0cm 0.5cm 1.0cm},clip,width=\linewidth]{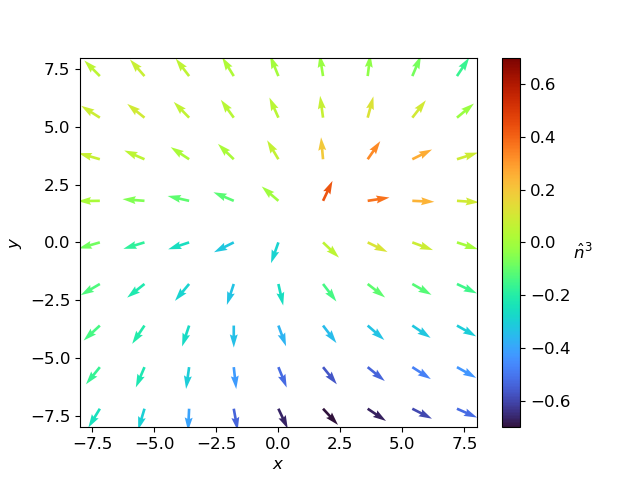}
    }\hfill
    \caption{The structure of the bilinear vectors, $R^a$ (top) and $n^a$ (bottom), surrounding a string that forms in simulations with random initial conditions. The vectors shown on the plots all have the same magnitude because they are the normalised 2-vectors formed from the $1$ and $2$ components of $n^a$ and $R^a$. The colour of each vector represents the $3$ component of the normalised full 3-vector. Both vector fields have also been subjected to a global rotation to make the visualisation of the structure more apparent.}
    \label{fig: bilinear vectors}
\end{figure}

We have performed simulations of the $\text{U(1)}_{\text{PQ}}$ symmetric 2HDM, in the global limit, from random initial conditions and in figure \ref{fig: Rp} we present evidence of neutral vacuum violation localised to the topological defects which has structure in the bilinear vectors that matches that of our solutions, as shown in figure \ref{fig: bilinear vectors}. These simulations have been performed in 2D to simplify the geometry so that the winding of the string is always simply in the $x-y$ plane. The parameters of the simulation are $\epsilon = 0.35$, $\delta = 0.5$ and $\alpha=\beta = \pi/4$ and we use $2048$ grid points in both directions with a lattice spacing of $0.9$ and a time-step of $0.2$. With $M_h^2=1$ and $v_{\text{SM}}^2=1$ fixed by rescaling, this corresponds to a simulation with $\mu_1^2=\mu_2^2=0.5$, $\lambda_1=\lambda_2=0.625$, $\lambda_3=0.995$ and $\lambda_4 = -0.245$.

The objects that form in these simulations have features that lead us to believe that they are very similar to the string solutions we have found. We see in figure \ref{fig: Rp} that neutral vacuum violation occurs in the vast majority, but not all, of the cores of the strings, which is as we expected from our mass analysis since $\epsilon<\delta$ and $\alpha=\beta$. We performed another simulation with the values of $\epsilon$ and $\delta$ interchanged in which strings form but neutral vacuum violation in the core is significantly more rare, as expected. 

Secondly and perhaps more convincingly, figure \ref{fig: bilinear vectors} shows that there is a winding in the $1$ and $2$ components of both $R^a$ and $n^a$ around the string. We also colour the vectors with the third component of the vector which shows only small contributions from $\hat{R}^3$. Since this is the global limit, the current term in the gradient energy leads to a rotation (as a function of the radius) between $\hat{n}^3$ and the other two components, which would be absorbed into the $W_r^2$ gauge field component for our

\noindent solutions. It is for this reason that there is also evidence of a winding around the string in the $\hat{n}^3$ component, which is less pronounced at greater distances from the string.

We have performed similar simulations for the case of monopoles and these are presented in our earlier work on global monopoles \cite{BATTYE2023138091}. Taken together with these simulations of vortices, they show that the solutions we have found are the topological solutions which are of most physical relevance.

\bibliographystyle{apsrev4-1}
\bibliography{supp_refs.bib}